---------------------------------------------------------------

\documentstyle[12pt]{article}
\begin{document}
\centerline{\bf SEMICLASSICAL TREATMENT OF THE DIRAC SEA CONTRIBUTION}
\centerline{\bf FOR FINITE NUCLEI }
\vspace{1cm}

\centerline{ J. Caro, E. Ruiz Arriola and L.L. Salcedo}
\vskip0.5cm
\centerline {Departamento de F\'{\i}sica Moderna, Universidad de Granada}
\centerline {E-18071 Granada, Spain }
\vskip 1.6cm

\centerline{\bf ABSTRACT}
\vskip 18pt
\vbox{
\narrower{
Dirac sea corrections for bulk properties of finite nuclei are
computed within a self-consistent scheme in the $\sigma$-$\omega$
model. The valence part is treated in the Hartree approximation
whereas the sea contribution is evaluated semiclassically up to fourth
order in $\hbar$.
Numerically, we find a quick convergence of the semiclassical
expansion; the fourth order contributing much less than one percent to
the binding energy per nucleon.
}}
\vskip0.6cm

\vskip0.5cm

\noindent
{\sc Preprint UG-DFM-5/96}


\vfill
\eject

\null\medskip
\section*{1 Introduction}
The quantum field theoretical approach to Nuclear Physics
seems to be the only practical device known up to now to include
relativistic corrections in the description of atomic nuclei in a
systematic way \cite{Wa74, Ho81, Se86, Re89, Se92}.
In most works the description is
limited to the valence part within a mean field Hartree
approximation. Much less work has been devoted to the study of the
Dirac sea of nucleons which
should be included in principle to preserve the unitarity of the theory.
The Dirac sea has been mostly considered in order to describe nuclear matter
\cite{Se86} or finite nuclei within the local density approximation up to
two loops \cite{Ta92}.
On the other hand, even at the lowest order of the loop expansion,
the physical relevance of the
single particle negative energy nucleon states in finite nuclei
is not fully understood. The explicit inclusion of the sea presents
both numerical and conceptual problems. A numerical computation of the
Dirac sea energy would require the diagonalization of all the continuum
states of the single particle Hamiltonian and subsequent renormalization.
A more serious problem is represented by the appearance of short
distance tachyonic
ghosts in the meson propagators at the one loop level which makes the
mean field vacuum state~\cite{Pe86,Co87} and nuclear matter~\cite{FH88}
unstable against formation of
translationally non-invariant configurations of size of the order of
0.2~fm.
In the relativistic approach to Nuclear Physics,
a prescription has recently been suggested to deal with the instability
problem based on the K\"allen-Lehmann representation of two-point Green's
functions which, so far, has only been applied to the study of nuclear
matter and the corresponding local density approximation \cite{Ta91,Ta92}.
In addition,
the method presents ambiguities \cite{Bo61} and its extension to arbitrary
Green's
functions is not known.

A way out of the above mentioned difficulties, which applies
specifically to the mean field approximation, is to treat the Dirac
sea assuming that the mesonic mean fields are smooth functions in
coordinate space, thus allowing for a derivative or semiclassical
expansion of the Dirac sea contribution to the energy, but keeping
the shell structure in the valence part. Indeed, the
problem of the high momentum instability of the sea
is circumvented since only low
Euclidean momenta are probed by the semiclassical expansion. Moreover,
the numerical problems related to the continuum states or, equivalently,
the non-local nature of the fermionic determinant, are drastically
reduced to the study of an analytically given local functional of the
mesonic fields and their lower derivatives.
In fact, such an approach has been applied to compute the
effect of the Dirac sea up to second order in $\hbar$~\cite{Pe86,FP90}.
There, it was found that such effects cannot be considered negligible.
Obviously,
there arises the question whether the higher order terms in the expansion can
be safely neglected. There is, however, an additional subtlety, namely,
even if the semiclassical expansion in powers of
$\hbar$ turns out to be numerically convergent, there still could
appear finite corrections of the type $\exp(-M_N R c/\hbar)$ which might
be numerically significant as compared to higher order $\hbar$
corrections. Such terms are beyond a semiclassical expansion and
represent a sort of shell effects which, for the valence part,
are known to be comparable to the second order $\hbar$
corrections~\cite{Ca96}.
Finally, an additional advantage of applying the semiclassical
expansion to the Dirac sea lies in the absence of turning point
divergences. In contrast, in the valence part these divergences take
place and become a serious problem beyond second order, particularly
in the formulation of a self-consistent approach.

In the present letter, we undertake the
calculation of the fourth order term of the Dirac sea contribution to
the binding energy per nucleon as well as other relevant observables.
This is done in a combined self-consistent Hartree treatment of
the positive energy single particle states and a semiclassical
expansion of the negative energy states.
For simplicity, we do so for the $\sigma$-$\omega$ model, although we
expect our considerations to hold in more realistic versions including,
e.g., $\rho$-meson as well as Coulomb interaction.

\section*{2 Field theoretical model}
The $\sigma$-$\omega$ model is characterized by the following
Lagrangian density~\cite{Wa74}
\begin{eqnarray}
{\cal L} & = &  \overline\Psi \left[ \gamma_\mu ( i \partial^\mu - g_v V^\mu)
- (M - g_s \phi) \right] \Psi + {1\over 2}\, (\partial_\mu \phi
\partial^\mu \phi - m_s^2 \, \phi^2) +  \nonumber \\
 & & \quad - {1\over 4} \, F_{\mu \nu} F^{\mu \nu}
+ {1\over 2} \, m_v^2 V_\mu V^\mu\,,
\label{eq1}
\end{eqnarray}
where $\Psi$ is the isospinor nucleon field, $\phi$ the scalar field,
$V_\mu$ the $\omega$-meson field and
$F_{\mu\nu} =\partial_\mu V_\nu-\partial_\nu V_\mu$.
In the former expression the
necessary counterterms required by renormalization are implicitly included.
In the time-independent mean field Hartree approximation the meson
fields are replaced by classical c-numbers whereas the fermionic state
is given by a Slater determinant consisting of valence plus sea single
particle states. The total mean field energy is given by
$E[\phi,V_0] = E^{\rm val} + E^{\rm sea} + E_B$, where
\begin{eqnarray}
E^{\rm val} &=& \sum_n E_n^{\rm val}\,,\qquad
E^{\rm sea} = \sum_n E_n^{\rm sea} \nonumber \\
E_B &=& {1\over 2}\int d^3x \, \left[ (\nabla \phi)^2 + m_s
\phi^2 - (\nabla V_0)^2 - m_v V_0^2 \right]\,.
\label{eq2}
\end{eqnarray}
and the single particle valence and sea orbitals depend functionally
on the mesonic fields $\phi$ and $V_0$ through the Dirac equation
\begin{equation}
\left[ -i {\bf \alpha} \cdot \nabla + g_v V_0({\bf r}) +
 \beta (M - g_s \phi({\bf r}))
\right] \psi_n({\bf r}) = E_n \,\psi_n({\bf r})\,.
\label{eq3}
\end{equation}
Notice that in the nuclear ground state of spherical nuclei,
the only case to be considered in this letter,
the spatial components of the $\omega$-meson field vanish \cite{Se92}.

To perform in practice the
derivative expansion we borrow from refs.~\cite{Ru93,Ca96}, where the single
particle level density $\rho(E)$ is obtained to fourth order in
$\hbar$ for an arbitrary space dimension $D$.
The sea energy can be deduced from eqs.~(3) and (18) of ref.~\cite{Ru93}
by means of the relation
\begin{equation}
E^{\rm sea} = \int_{-\infty}^0 dE\,E\,\rho(E)\,.
\label{eq4}
\end{equation}
The sea energy is ultraviolet divergent and requires a suitable
renormalization. The standard dimensional regularization can be applied with
$D=3-2\epsilon$. Since we have in mind a semiclassical expansion of
the sea energy we specify the renormalization conditions
at zero momentum as done in ref.~\cite{Se92}. This results in the following
expression for the renormalized sea energy to fourth order in the
gradients of the mesonic fields
\begin{eqnarray}
E^{\rm sea}_0 & = & -{\gamma\over 16\pi^2} M^4 \int d^3{\bf r} \,
\Biggl\{ \Biggr.
\left({\Phi\over M}\right)^4 \log {\Phi\over M}
+ {g_s\phi\over M} - {7\over 2} \left({g_s \phi\over M}\right)^2
\nonumber \\
 & & \qquad
+ {13\over 3} \left({g_s \phi\over M}\right)^3
- {25\over 12} \left({g_s \phi\over M}\right)^4
 \Biggl. \Biggr\} \nonumber\\
E^{\rm sea}_2 & = & {\gamma \over 16 \pi^2} \int d^3 {\bf r} \,
\Biggl\{ {2 \over 3} \log{\Phi\over M} (\nabla V)^2
- \log{\Phi\over M} (\nabla \Phi)^2 \Biggr\} \nonumber\\
E^{\rm sea}_4 & = & {\gamma\over 5760 \pi^2} \int d^3 {\bf r} \,
\Biggl\{ \Biggr. -11\,\Phi^{-4} (\nabla \Phi)^4
- 22\,\Phi^{-4} (\nabla V)^2(\nabla \Phi)^2 \nonumber\\
 & & \quad
+ 44 \, \Phi^{-4} \bigl( (\nabla_i \Phi) (\nabla_i V) \bigr)^2
- 44 \, \Phi^{-3} \bigl( (\nabla_i \Phi) (\nabla_i V) \bigr) (\nabla^2 V)
\nonumber\\
 & & \quad
- 8 \, \Phi^{-4} (\nabla V)^4
+ 22 \, \Phi^{-3} (\nabla^2 \Phi) (\nabla \Phi)^2
+ 14 \, \Phi^{-3} (\nabla V)^2 (\nabla^2 \Phi)
\nonumber\\
 & & \quad
- 18 \, \Phi^{-2} (\nabla^2 \Phi)^2 + 24 \, \Phi^{-2} (\nabla^2 V)^2
\Biggl. \Biggr\}\,.
\label{eq5}
\end{eqnarray}
Here, $V=g_vV_0$,~$\Phi=M-g_s\phi$ and $\gamma$ is the spin and
isospin degeneracy of the nucleon, i.e., four if there are two nucleon species.
The first two terms have been known for some time, see e.g.~\cite{Se92}.
The fourth order term is new and turns out to be ultraviolet convergent.
This was expected in advance since dimensional counting shows that
there is a correspondence between the number of gradients
and the superficial degree of divergence of the
corresponding momentum integral.

\section*{3 Numerical results}
The mean field equations of motion are obtained, as usual, by minimizing the
total mean field energy functional, $E[\phi,V_0]$ (see definition
above eq.~(\ref{eq2})), with respect to arbitrary
variations of $\phi({\bf r})$ and $V_0({\bf r})$. This yields
eq.~(\ref{eq3}) together with
\begin{eqnarray}
(\nabla^2-m_s^2)\phi({\bf r}) &=&-g_s(\rho_s^{\rm val}({\bf r}) +
 \rho_s^{\rm sea}({\bf r})) \nonumber \\
(\nabla^2-m_v^2)V_0({\bf r}) &=&-g_v(\rho_v^{\rm val}({\bf r}) +
 \rho_v^{\rm sea}({\bf r}))
\label{eq6}
\end{eqnarray}
where we have explicitly separated the valence and sea contributions
to the scalar and baryonic densities, $\rho_s=\langle\overline\Psi\Psi\rangle$
and $\rho_v=\langle\Psi^\dagger\Psi\rangle$, respectively.
For both densities, the valence part is just the corresponding
sum over valence orbitals. On the other hand, the sea densities can be
obtained as
\begin{equation}
\rho^s({\bf r}) = -{1\over g_s}{\delta E^{\rm sea}\over\delta
\phi({\bf r})}\,,\qquad
\rho^v({\bf r}) = {1\over g_v}{\delta E^{\rm sea}\over\delta
V_0({\bf r})}\,.
\label{eq7}
\end{equation}
In turn, the expansion for $E^{\rm sea}$ given in eq.~(\ref{eq5}) provides a
similar semiclassical expansion for the scalar and baryonic densities.
We will consider closed-shell nuclei for which the equations admit spherically
symmetric solutions. These equations must be solved self-consistently.
This poses no particular problems when the sea is included to zeroth
or second orders since they give rise to second order differential
equations.
The fourth order case is qualitatively different since it
requires to solve fourth order differential equations and consequently
further boundary conditions need to be specified. Instead of doing so,
the fourth order sea contributions are treated perturbatively,
that is, we consider the full self-consistent solution to second order
and use the resulting meson fields to compute the fourth order
sea corrections to the binding energy and the nuclear radius.
Obviously, due to the stationarity of the energy functional to second
order, this procedure reproduces the correct fourth order correction to the
energy up to higher order terms. As it will be shown below, the fourth
order corrections turn out to be small, thus justifying {\it a posteriori}
this procedure.

In Table~\ref{tabA} we present
results for the binding energy per nucleon and mean squared charge
radius for several closed-shell nuclei calculated in various
approximations. In all cases, the parameters of the Lagrangian, namely,
$g_s$, $g_v$, $m_s$ and $m_v$, are adjusted to reproduce the
$\omega$-meson mass, the saturation density and binding energy per
nucleon for nuclear matter and the mean squared charge radius of
$^{40}$Ca ~\cite{Ho81,Se92}.
The corresponding numerical values are given in Table~\ref{tabB}.
In both tables the entry ``no sea'' stands for the customary
valence Hartree approximation, where the effects due to the Dirac sea
are fully neglected. 0th- and 2nd-order represent the results of
successive inclusion of the Dirac sea keeping the corresponding orders
in the expression for the sea energy and sea densities in
eqs.~(\ref{eq5}) and (\ref{eq7}).
The experimental values are given only for illustration, since the
$\sigma$-$\omega$ model must be supplemented with
additional degrees of freedom, such as
the $\rho$-meson and Coulomb interaction, to be realistic~\cite{Se86,Se92}.
{}From
Table~\ref{tabA}, one can see that the Dirac sea corrections are not globally
small and hence cannot be neglected. The effect becomes more dramatic
on the fixing of the parameters, Table~\ref{tabB}.
Finally, in Table~\ref{tabC} we present our results for the full calculation
where the Dirac sea is described semiclassically up to fourth order in
$\hbar$. The various contributions to minus the total binding
energy per nucleon are displayed, namely, valence kinetic energy,
valence potential energy, minus sea potential energy and total sea energy.
Their total sum yields minus the binding energy per nucleon.

In Table~\ref{tabA}, the results for the fourth order have been
intentionally omitted since they do not further modify the presented
second order results. In fact, in all cases considered, i.e.,
closed-shell nuclei, the fourth order contributions are very small as
they do not significantly influence neither the bulk properties nor
the parameters of the model. The smallness of the correction does not
follow from big cancelations of any kind; each term in eq.~(\ref{eq5})
turns out to be uniformly small in the integration region.
This is an important result for it
indicates that the semiclassical expansion converges much faster than
one might naively expect. Indeed, the fourth order sea correction to
the binding energy per nucleon can be estimated, from eq.~(\ref{eq5}),
to be of the order of
$1/(RA)$, $R$ being the nuclear radius and $A$ the mass number. In
the case of calcium it corresponds to a correction of about 1~MeV. The
fact that this number turns out to be much smaller is a direct
consequence of the overall numerical dimensionless coefficient, whose
value requires an explicit calculation, as done here. The higher
orders are not expected to be relevant. For instance, the sixth order can
be estimated to be suppressed by a further factor $1/(UR)^2$, $U$ being the
depth of either the scalar or the vector potentials, i.e., about
one percent of the fourth order correction for calcium.

As already mentioned in the introduction, a word of caution is in order.
The fast numerical convergence of the semiclassical expansion for the sea
contribution to the mean field energy does not necessarily
imply that it converges to the exact Hartree result. The
expansion might be only asymptotic or it might be
convergent but not to the exact result.
Quite generally, non analytical terms in $\hbar$,
which are beyond a semiclassical approach, are expected to occur in
global properties.
Actually, such contributions can be shown to appear in simple non
relativistic quantum mechanical models~\cite{Ca96b}.
The existence of these terms must be kept in mind if the semiclassical
results are to be compared with an exact calculation, since they might
introduce systematic deviations. Such deviations seem to have been
observed in the quark chiral soliton model~\cite{Me88}.

\section*{4 Conclusions}
We summarize our points. The Dirac sea corrections
to the mass and size of finite nuclei have been computed in a
semiclassical self-consistent treatment of the $\sigma$-$\omega$
model, but taking into account the shell structure of the valence part
by means of a Hartree approximation. In this combined self-consistent
approach, we have
investigated the numerical convergence of the semiclassical expansion
up to fourth order in $\hbar$ for the considered bulk properties.
Although the Dirac sea as a whole cannot be considered negligible, the
overall effect becomes smaller after the necessary readjustment of the
parameters in the model. Furthermore, a fair description of the sea
can be achieved by including just the zeroth and second order
semiclassical terms. We emphasize the non triviality of this result
since a simple dimensional estimate allows much larger corrections
than the ones actually found at fourth order.
Finally, the present
calculations can be regarded as a previous step to an exact mean field
description of the polarization of the Dirac sea.

\section*{Acknowledgments}

J. Caro acknowledges the Spanish M.E.C. for a grant.
This work has been partially supported by the DGICYT under contract
PB92-0927 and the Junta de Andaluc\'\i a (Spain).

\newpage

\section*{Table Captions}
\begin{enumerate}
\item Binding energy per nucleon (in MeV) and mean squared charge
radius (in fm) for closed-shell nuclei in successive calculational schemes.
The ``no sea'' column stands for the purely valence calculation. The
``sea 0th'' and ``sea 2nd'' columns refer to inclusion of the sea
through a semiclassical expansion. In all cases the parameters have
been readjusted as explained in the main text.
Experimental values are only given as orientative reference.
\label{tabA}

\item Parameters of the $\sigma$-$\omega$ model adjusted to
fit the mean squared charge radius of $^{40}$Ca and nuclear matter
saturation density and binding energy (see ref.~\cite{Se86} for the
definition of the parameters). The nucleon and $\omega$ meson
masses are fixed to 939~MeV and 783~MeV respectively.
The same calculational
schemes as in the Table~\ref{tabA} are considered. The scalar meson
mass is given in MeV.
\label{tabB}

\item In different columns, the valence kinetic energy,
the valence potential energy, minus the
sea potential energy and the total sea energy per nucleon are displayed for the
full fourth order semiclassical approximation to the Dirac sea.
Their total sum yields minus the binding energy per nucleon.
\label{tabC}
\end{enumerate}

\newpage
\begin{figure}
\begin{center}
\begin{tabular}{|c|*{2}{r@{.}l|}*{2}{r@{.}l|}*{2}{r@{.}l|}*{2}{r@{.}l|}}
\cline{2-17}
\multicolumn{1}{c}{ }
        & \multicolumn{4}{|c|}{No sea}
        & \multicolumn{4}{|c|}{Sea 0th}
        & \multicolumn{4}{|c|}{Sea 2nd}
        & \multicolumn{4}{|c|}{Exp.}
\\ \hline
$^A_Z$X & \multicolumn{2}{|c|}{B/A}
        & \multicolumn{2}{|c|}{m.s.c.r.}
        & \multicolumn{2}{|c|}{B/A}
        & \multicolumn{2}{|c|}{m.s.c.r.}
        & \multicolumn{2}{|c|}{B/A}
        & \multicolumn{2}{|c|}{m.s.c.r.}
        & \multicolumn{2}{|c|}{B/A}
        & \multicolumn{2}{|c|}{m.s.c.r.}
\\ \hline
$^{40}_{20}$Ca  & 6&28   & 3&48$^{*}$           &
		  6&00   & 3&48$^{*}$           &
		  6&33   & 3&48$^{*}$           &
                  8&55   & 3&48                 \\
$^{48}_{20}$Ca  & 6&52   & 3&47\phantom{$^{*}$} &
		  6&03   & 3&51                 &
		  6&34   & 3&52                 &
                  8&67   & 3&47                 \\
$^{56}_{28}$Ni  & 7&24   & 3&72\phantom{$^{*}$} &
		  6&51   & 3&79                 &
		  6&80   & 3&79                 &
                  8&64   & \multicolumn{2}{|c|}{} \\
$^{90}_{40}$Zr  & 8&36   & 4&22\phantom{$^{*}$} &
		  7&99   & 4&23                 &
		  8&22   & 4&25                 &
                  8&71   & 4&27                 \\
$^{132}_{50}$Sn & 8&81   & 4&60\phantom{$^{*}$} &
		  8&43   & 4&66                 &
		  8&62   & 4&68                 &
                  8&36   & \multicolumn{2}{|c|}{}\\
$^{208}_{\phantom{2}82}$Pb & 9&84   & 5&35\phantom{$^{*}$} &
			     9&55   & 5&39                 &
	       		     9&70   & 5&41                 &
                             7&87   & 5&50                 \\
\hline
\end{tabular}
\vspace{.5cm}

Table \ref{tabA}
\end{center}
\end{figure}

\begin{figure}
$$
\begin{array}{|c|*{5}{r@{.}l|}}
\cline{2-11}
\multicolumn{1}{c}{ }
       & \multicolumn{2}{|c|}{C_s^2}   & \multicolumn{2}{|c|}{m_s}
&\multicolumn{2}{|c|}{g_s}      &  \multicolumn{2}{|c|}{C_v^2}   &
\multicolumn{2}{|c|}{g_v}    \\ \hline
{\rm No sea}
       & 357&741   &   449&93    &  9&06283   &  274&106   &  13&8056  \\
{\rm Sea 0th}
       & 227&840 &   368&37   &  5&92153  & 147&524  &  10&1281  \\
{\rm Sea 2nd}
       & 227&840 &   406&6    &  6&53607  & 147&524  &  10&1281  \\
\hline
\end{array}
$$
\vspace{.3cm}

\centering{Table \ref{tabB}}
\end{figure}

\begin{figure}
\begin{center}
\begin{tabular}{|c|*{5}{r@{.}l|}r@{.}l@{$\cdot$}l|}
\hline
$^A_Z$X
  & \multicolumn{2}{|c|}{K$^{\rm val}$}
  & \multicolumn{2}{|c|}{\rule{0cm}{.65cm}$-{1 \over 2} \rho^{\rm val}$ U}
  & \multicolumn{2}{|c|}{$ {1 \over 2} \rho^{\rm sea}$ U}
  & \multicolumn{2}{|c|}{E$_0^{\rm sea}$}
  & \multicolumn{2}{|c|}{E$_2^{\rm sea}$}
  & \multicolumn{3}{|c|}{E$_4^{\rm sea}$}
\\ \hline
$^{40}_{20}$Ca  &  14&6  & $-$18&0  &  $-$5&29  & 1&68 & 0&681 &
$-$2&6&10$^{-4}$\\
$^{48}_{20}$Ca  &  15&6  & $-$18&8  &  $-$5&61  & 1&82 & 0&670 &
8&3&10$^{-5}$\\
$^{56}_{28}$Ni  &  16&1  & $-$19&5  &  $-$6&11  & 2&01 & 0&679 &
6&7&10$^{-4}$\\
$^{90}_{40}$Zr  &  16&1  & $-$20&3  &  $-$7&02  & 2&42 & 0&586 &
$-$1&4&10$^{-5}$\\
$^{132}_{50}$Sn &  16&7  & $-$21&0  &  $-$7&52  & 2&66 & 0&526 &
$-$1&0&10$^{-4}$\\
$^{208}_{\phantom{2}82}$Pb
                &  16&8  & $-$21&7  &  $-$8&31  & 3&01 & 0&455 &
$-$1&9&10$^{-4}$\\
\hline
\end{tabular}
\vspace{.5cm}

Table \ref{tabC}
\end{center}
\end{figure}
\end{document}